\documentclass[entropy,article,accept,moreauthors,pdftex,10pt,a4paper]{mdpi} 

\usepackage{amsmath}
\usepackage{amsfonts}
\usepackage{graphicx}
\usepackage{caption}
\usepackage{subcaption}

 \usepackage{amssymb}

\usepackage{epstopdf}
\usepackage{epsfig}

\usepackage{hyperref}

\usepackage{color}
  
\usepackage{microtype} 

\usepackage{graphicx}
\makeatletter      
                    \renewcommand{\maketag@@@}[1]{\hbox{\m@th\normalsize\normalfont#1}}%      
                    \makeatother

\firstpage{1} 
\articlenumber{438}
\doinum{10.3390/e18120438}
\pubvolume{18}
\pubyear{2016}
\copyrightyear{2016}
\externaleditor{Academic Editor: David Kubiznak}
\history{Received: 28 October 2016; Accepted: 1 December 2016; Published: 8 December 2016}
\graphicspath{{./figures/}}
\usepackage{soul}

\Title{Static Einstein--Maxwell Magnetic Solitons and Black Holes in an Odd Dimensional AdS Spacetime}

\Author{Jose Luis Bl\'azquez-Salcedo $^{1,}$*, Jutta Kunz $^{1}$ , Francisco Navarro-L\'erida $^{2}$ and Eugen Radu $^{3}$} 
\AuthorNames{Jose Luis Bl\'azquez-Salcedo, Jutta Kunz, Francisco Navarro-L\'erida and Eugen Radu}

\address{%
$^{1}$ \quad Institut f\"ur  Physik, Universit\"at Oldenburg, Postfach 2503, 
  D-26111 Oldenburg, Germany; jutta.kunz@uni-oldenburg.de\\
$^{2}$ \quad Departamento~de F\'{\i}sica At\'omica, Molecular y Nuclear, Ciencias F\'{\i}sicas 
 Universidad Complutense de Madrid, E-28040 Madrid, Spain; fnavarro@fis.ucm.es\\
 $^{3}$ \quad Departamento de F\'\i sica da Universidade de Aveiro and CIDMA, Campus de Santiago, 3810-183 Aveiro, Portugal; eugen.radu@ua.pt}

\corres{Correspondence: jose.blazquez.salcedo@uni-oldenburg.de; Tel.: +49-441-798-3469}

\abstract{ We construct a new class of Einstein--Maxwell static  solutions with a magnetic field
in $D$-dimensions (with $D\geq 5$ an odd number), approaching at infinity a globally Anti-de Sitter (AdS) spacetime. In addition to the mass,  the new solutions possess an extra-parameter associated with a non-zero magnitude of the magnetic potential at infinity. Some of the black holes possess a non-trivial zero-horizon size limit, which corresponds to a solitonic deformation of the AdS background.}


\begin{document}

%%%%%%%%%%%%%%%%%%%%%%%%%%%%%%%%%%%%%%%%%%
%% Only for the journal Gels: Please place the Experimental Section after the Conclusions

%%%%%%%%%%%%%%%%%%%%%%%%%%%%%%%%%%%%%%%%%%

\section{Introduction}

A simple scaling argument, going back at least to Coleman \cite{Coleman:1977hd} and Deser \cite{Deser:1983fk}, shows that no gauge field solitons exist in a four-dimensional 
Minkowski spacetime background.
Moreover, for a Maxwell field, this also holds when taking into account the backreaction on the geometry \cite{Shiromizu:2012hb} (and keeping the assumption of asymptotic flatness). 
The situation is different, however, for a (globally) Anti-de Sitter (AdS) background.
An early indication in this direction came from the discovery in \cite{BoutalebJoutei:1979va}
of an exact solution of the Yang--Mills (YM) equations in an AdS spacetime (note 
that only the dS version of the solution was considered in \cite{BoutalebJoutei:1979va}, while
the AdS interpretation has been given much later in \cite{VanderBij:2001ah}).
%Footnote is not preferred. We refer to it in the main text. Please confirm if this is okay.
This solution is regular everywhere, has finite mass and describes a unit charge magnetic monopole. Different from the well known (flat spacetime) 't Hooft--Polyakov
monopole \cite{'tHooft:1974qc,Polyakov:1974ek}, it exists without a Higgs field, being supported by 
a nonzero value of the cosmological constant $\Lambda$.

In fact, as found in \cite{Hosotani:2001iz}, the configuration in \cite{BoutalebJoutei:1979va}
is a particular member of a family of solutions, which are classified by an (arbitrary)  
magnetic charge. As expected, these solitons survive when taking into account the gravity effects
\cite{Bjoraker:2000qd} implying the existence of black holes (BHs) with non-Abelian hair~\cite{Winstanley:1998sn}. The gravitating AdS solutions possess a variety of interesting features that strongly contrast with those of the asymptotically flat spacetime counterparts  
\cite{Volkov:1989fi,Kuenzle:1990is,Bizon:1990sr, Volkov:1998cc} (for example, some of them are stable).
As such, the study of the Einstein--Yang--Mills-AdS (EYM-AdS) solitons and BHs has been an active field of research over the last decade (for a review, see \cite{Winstanley:2008ac}).

In hindsight, the existence of such solutions is not a surprise, and can be attributed to the fact that the AdS spacetime effectively acts like a 'box'. Then, the field modes, which, for $\Lambda=0$, are divergent at infinity, get regularized. In addition, some of the features can be attributed to the non-linearities of the model.

In a rather unexpected development, it has been recently realized \cite{Herdeiro:2015vaa, Costa:2015gol, Herdeiro:2016xnp, Herdeiro:2016plq} that even an Abelian gauge field possesses solitonic solutions in the AdS background. Such solutions exist for $any$ term in the multipole 
expansion, except the lowest one. In addition, they can be promoted to gravitating solitons
of the full Einstein--Maxwell (EM) theory when including the backreaction, without destroying the AdS asymptotics. Moreover, placing a horizon inside these solitons results in static EM BHs, which are very different from the well known Reissner--Nordstr\"om-AdS solution. 

The results above concern the (most familiar) case $D=4$. However, it is worth inquiring about the situation in more than four dimensions. Since the  ``boxing'' feature is not specific to AdS$_4$ spacetime, this suggests that similar results should also be found for $D>4$ dimensions.
Indeed, this is the case for non-Abelian fields, the solutions in \cite{BoutalebJoutei:1979va, Hosotani:2001iz} possessing higher dimensional generalizations with many similar properties 
\cite{Okuyama:2002mh, Radu:2005mj}. Thus, it is then natural to expect that the same holds for a Maxwell field. 

The main purpose of this work is to report the results of a preliminary investigation in this direction. We shall focus on a particular class of static solutions with  magnetic fields only,
approaching at infinity an odd-dimensional (globally) AdS background. This makes it possible to introduce an ansatz that factorizes the angular dependence, such that the EM system of equations reduces to a codimension-1 problem (although the system is $not$ spherically symmetric). These aspects are discussed in Sections 2 and 3 of this work.

The emerging picture for $D=5$, $7$, and $9$ shows some similarities with the $D=4$ results in \cite{Herdeiro:2015vaa, Costa:2015gol, Herdeiro:2016xnp, Herdeiro:2016plq}. For example, in the probe limit, one again finds U(1) solitons. In addition, as discussed in Section 4, these configurations possess gravitating generalizations, including BHs with a nontrivial magnetic field.  However, some properties are different. For example, due to the slow decay of the magnetic fields, in higher dimensions, one has to supplement the action with a boundary matter counterterm, despite the spacetime being asymptotically AdS. 

%%%%%%%%%%%%%%%%%%%%%%%%%%%%%%%%%%%%%%%%%%%%%%%%%%%%%%%%%%%%%%%%%%%
 \section{Magnetic Fields in an Odd-Dimensional AdS Spacetime}
%%%%%%%%%%%%%%%%%%%%%%%%%%%%%%%%%%%%%%%%%%%%%%%%%%%%%%%%%%%%%%%%%%%
Before approaching the issue of EM solutions,
it is useful to consider first the probe limit, i.e.,
a~U(1) field in a fixed AdS spacetime.
Restricting to $D=2k+3$ dimensions (with $k\geq 1$ as such $D\geq 5$),
we consider a general metric ansatz with
\begin{eqnarray}
\label{metric-in}
ds^2=\frac{dr^2}{N(r)}+r^2 d\Omega_{2k+1}^2-N(r)dt^2.
\end{eqnarray}
In our approach,
 the metric of the odd-dimensional 
(round) sphere
is written
as an $S^1$ fibration over the complex projective
space $\mathbb C \mathbb P^k$,
\begin{eqnarray}
\label{s1}
d\Omega_{2k+1}^2=(d\psi+{\cal A})^2+d\Sigma_k^2,
\end{eqnarray}
where $d\Sigma_k^2$
is the Fubini-Study metric on the unit $\mathbb C \mathbb P^k$ space and
%Should this be an en-dash instead of a hyphen?
${\cal A}=A_i dx^i $ is its K\"ahler form. 
%Footnote is not preferred. We refer to it in the main text. Please confirm if this is okay.
 The fibre is parameterized by
the coordinate $\psi$, which has period $2\pi$.
A simple explicit form for Equation (\ref{s1})
%Pls add a Label here or change ``Lable'' into ``Equation''???
is found by introducing
$k+1$
complex coordinates $z_i$ (with $\displaystyle\sum_{i}^{k+1} z_i  \bar{z}_i=1$),
 such that
$d\Omega_{2k+1}^2=\displaystyle\sum_{i}dz_id\bar{z}_i$.
Then, one can take (see e.g.,  \cite{Stotyn:2013yka}):
%\begin{eqnarray}
$
z_i=   e^{i(\psi+\phi_i)}\cos\theta_i\displaystyle\prod_{j<i}\sin\theta_j,
$
for 
$i=1,\dots,k,$
and
$
%z_{k+1}= e^{i\psi} \prod_{j=1}^{\frac{n-1}{2}}\sin\theta_j.
z_{k+1}= e^{i\psi} \prod_{j=1}^{k}\sin\theta_j.
$
%\end{eqnarray}
Note that the coordinates $\phi_i$ have period $2\pi$, while the $\theta_i$ have period $\pi/2$.
The~corresponding expression of the  K\"ahler form ${\cal A}$ is
$
{\cal A}={\cal A}_i dx^i=\sum_{i=1}^{k}{\cos^2\theta_i\left[\prod_{j<i}\sin^2\theta_j\right]d\phi_i}~$. 

A general study of the Maxwell equations in the background (\ref{metric-in},\ref{s1}) 
(for a given $N(r)$)
is a complicated task.
However, the situation simplifies dramatically
for a gauge field ansatz 
%Footnote is not preferred. We refer to it in the main text or you can refer it in other places of the main text. Please confirm if this is okay.
:
\begin{eqnarray}
\label{U1}
B= B_i dx^i=a_\varphi(r)  \big(d\psi+ {\cal A} \big)
\end{eqnarray}
(with $B$ the gauge potential and the corresponding field strength tensor $F=dB$),
which factorizes the angular dependence.
This expression of the gauge potential is essentially
the magnetic truncation of the U(1) ansatz  used in 
\cite{Kunz:2006eh,Kunz:2007jq} in the study of Einstein--Maxwell(--Chern--Simons) rotating black holes in $D=2k+3$ dimensions.

This is a consistent ansatz, the Maxwell equations
$dF=0$
reducing to a single ordinary differential equation (ODE) for
 $a_\varphi(r)$:

\begin{eqnarray}
\label{eqA}
r''a_\varphi''+\left (\frac{rN'}{N}+D-4 
\right)r a_\varphi'-\frac{2(D-3)}{N}a_\varphi=0.
\end{eqnarray}
For a flat spacetime, $N(r)=1$, the above equation has the general 
solution
\begin{eqnarray}
\label{sol1}
a_\varphi(r)=\frac{c_1}{r^{D-3}}+c_m r^2
\end{eqnarray}
(with $c_1,c_m$ arbitrary constants),
which necessarily diverges at the origin or at infinity.
However,  
the~singularity at infinity is cured for an AdS background, 
where the metric function $N(r)$ in (\ref{metric-in}) is
\begin{eqnarray} 
\label{sol23}
N(r)=1+\frac{r^2}{L^2},
 \end{eqnarray}
with $L$ the AdS length scale.
In this case,  $ {c_1}/{r^{D-3}}$
is still a solution of Equation (\ref{eqA}),
 as in the flat spacetime limit.
However, the second independent solution 
is everywhere regular, in particular at $r = 0$,
and reads (with ${}_2F_1$
the hypergeometric function)

\vspace{-8pt}
 {\fontsize{9}{9}\selectfont\begin{eqnarray} 
\label{probe-gen}
 a_\varphi(r)=c_m \frac{D-3}{D-1}
 ~{}_2F_1\left(\frac{1}{2}(D-1), 1 ;  \frac{1}{2} (D+1) ;  - \frac{r^2}{L^2} \right) \frac{r^2}{L^2}.
 \end{eqnarray}}Here, the normalization has been chosen such that $a_\varphi(r)\to c_m$ as $r\to \infty$,
(with $c_m$ an arbitrary nonzero constant), while $a_{\varphi}\to c_m\frac{(D-3)r^2}{(D-1)L^2}$ as $r\to 0$.
More precisely, one finds

\vspace{-8pt}
 {\fontsize{9}{9}\selectfont\begin{eqnarray}
\label{probe-ex}
&&
\nonumber
 a_\varphi(r)=c_m  
\left[
1-\frac{L^2}{r^2}\log N(r)
\right], ~~~{\rm for}~~D=5,
\\
&&
a_\varphi(r)=c_m  
\left[
1-\frac{2L^2}{r^2}+\frac{2L^4}{r^4}\log N(r)
\right], ~~~{\rm for}~~D=7,
\\
&&
\nonumber
 a_\varphi(r)=c_m  
\left[
1-\frac{3L^2}{2r^2}+\frac{3L^4}{r^4}-\frac{3L^6}{r^6}\log N(r)
\right], ~~~{\rm for}~~D=9.
  \end{eqnarray}}

The profile of solutions together with their basic properties 
are similar to those discussed in Section 4 for gravitating
generalizations.

Rather similar results are found when considering instead
a spherically symmetric BH background.
Starting again with  the $\Lambda=0$ case,
we take 
\begin{eqnarray} 
\label{flat1}
N(r)=1-\bigg(\frac{r_H}{r} \bigg)^{D-3}~,
 \end{eqnarray}
such that the line element (\ref{metric-in}) 
corresponds to a Schwarzschild--Tangherlini BH
(with $r=r_H>0$ the event horizon radius).
In this case,
we notice the existence of the following exact solutions
of Equation~(\ref{eqA}) 
 
\vspace{-10pt}
 {\fontsize{9}{9}\selectfont
\begin{eqnarray}
\label{flat-d5}
\nonumber
&&a_\varphi(r)=c_m r^2+c_1
\left[
1+\left(\frac{r}{r_H}\right)^2\log \left(1-\left(\frac{r_H}{r}\right)^2 \right)
\right], ~~~
{\rm for}~~D=5,
\\
\label{flat-d7}
&&a_\varphi(r)=c_m r^2+c_1
\left[
2+\left(\frac{r}{r_H}\right)^2\log \left(\frac{1-(\frac{r_H}{r})^2}{1+(\frac{r_H}{r})^2} \right)
\right], ~~~~~
{\rm for}~~D=7,
\\
\nonumber
&&a_\varphi(r)=c_m r^2+c_1
\left[
6-2\sqrt{3}\left(\frac{r}{r_H}\right)^2\arctan\left(\frac{\sqrt{3}r_H^2}{2r^2+r_H^2}\right)
 \right.
\\
\nonumber
&&\left. \qquad \qquad \qquad \qquad \; \;  +\left(\frac{r}{r_H}\right)^2\log \left(\frac{(1-(\frac{r_H}{r})^2)^2}{1+(\frac{r_H}{r})^2+(\frac{r_H}{r})^4} \right)
\right], ~~~
{\rm for}~~D=9~.
\end{eqnarray}}Similar solutions can be constructed for higher $D$,
without it being possible to find the general pattern.
One can see that 
$a_\varphi(r)$ diverges at infinity 
or possesses a logarithmic singularity at the horizon,
a result which agrees with our intuition
based on the 'no-hair' conjecture.

\textls[-19]{The situation changes, however, when considering a Schwarzschild-AdS (SAdS)
background, with }
\begin{eqnarray}
\label{AdS1}
N(r)=1+\frac{r^2}{L^2}-\left(\frac{r_0}{r} \right)^{D-3} ,
\end{eqnarray}
with a horizon located at $r=r_H$, and
%\begin{eqnarray}
%\label{AdS11}
$
r_0=r_H\left(1+\frac{r_H^2}{L^2} \right)^{\frac{1}{D-3}}.
$
%\end{eqnarray}
Unfortunately, no exact solution could be constructed this time.
However, its numerical construction is straightforward, 
starting with the following near horizon expression
(with $a_\varphi^{(0)}$ some parameter)
{\fontsize{9}{9}\selectfont\begin{eqnarray}
\label{AdS2}
a_\varphi(r)=a_\varphi^{(0)}+\frac{2(D-3)a_\varphi^{(0)} }{r_H\left[D-3+(D-1)\frac{r_H^2}{L^2}\right]}(r-r_H)+\dots~.
\end{eqnarray}}
At infinity, the leading order terms of the solution are

\vspace{-10pt}
 {\fontsize{9}{9}\selectfont \begin{eqnarray}
\label{AdS12}
&&
a_\varphi(r)= c_m
    \left[
1+\log \bigg(\frac{L^2}{r^2} \bigg) \bigg(\frac{L}{r} \bigg)^2
-  \bigg(\frac{L}{r} \bigg)^4
+\frac{r_0^2}{3L^2}\log \bigg(\frac{L^2}{r^2} \bigg) \bigg(\frac{L}{r} \bigg)^6
+ \bigg(\frac{1}{2}+ \frac{2r_0^2}{9L^2}  \bigg) \bigg(\frac{L}{r} \bigg)^6
     \right]
		\\
		\nonumber
&&{~~~~~~~~~~~~~~~~~~~~~~~~~~~~~~~~~~~~~~~~~~~~~~~~~~~~~~~~~~~~~~~~~~~~~~~~~~~~}
+\mu\left[  \bigg(\frac{L}{r} \bigg)^2+\frac{r_0^2}{3L^2} \bigg(\frac{L}{r} \bigg)^6 \right]+\dots
,~~~~~~~~{\rm for}~~D=5,
 \\
		\nonumber
&&a_\varphi(r)= c_m
\left[
1-2\bigg(\frac{L}{r} \bigg)^2
-2\log \bigg(\frac{L^2}{r^2}\bigg)\bigg(\frac{L}{r} \bigg)^4
+2\bigg(\frac{L}{r} \bigg)^6
-\left(1+\frac{r_0^4}{2L^4}\right)\bigg(\frac{L}{r} \bigg)^8
\right]
 \\
\nonumber				
&&{~~~~~~~~~~~~~~~~~~~~~~~~~~~~~~~~~~~~~~~~~~~~~~~~~~~~~~~~~~~~~~~~~~~~~~~~~~~~~~~~~~~~~~~~~~~~~~~~~~~~~~~~~~~~~}+\mu \bigg(\frac{L}{r} \bigg)^4+\dots,~~~~{\rm for}~~D=7,
\end{eqnarray} }
and, for $D=9,$ 
 {\fontsize{9}{9}\selectfont\begin{eqnarray}
\label{AdS3}
		\nonumber
&&a_\varphi(r)= c_m
\left[
1-\frac{3}{2}\bigg(\frac{L}{r} \bigg)^2
+3\bigg(\frac{L}{r} \bigg)^4
+3\log \bigg(\frac{L^2}{r^2}\bigg)\bigg(\frac{L}{r} \bigg)^6
-3\bigg(\frac{L}{r} \bigg)^8
+\frac{3}{10}\left(5-\frac{r_0^6}{L^6}\right)\bigg(\frac{L}{r} \bigg)^{10}
\right]	
 \\
\nonumber				
&&{~~~~~~~~~~~~~~~~~~~~~~~~~~~~~~~~~~~~~~~~~~~~~~~~~~~~~~~~~~~~~~~~~~~~~~~~~~~~~~~~~~~~~~~~~~~~~~~~~~~~~~~~~~~~~~~~~~~~~~~~~~~~~~~~~~~}
+\mu \bigg(\frac{L}{r} \bigg)^6+\dots,
\end{eqnarray}}with $c_m$  and $\mu$  arbitrary constants
(note also the existence in the far field expression of terms induced by
the presence of the horizon).

%%%%%%%%%%%%%%%%%%%%%%%%%%%%%%%%%%%%%%%%%%%%%%%%%%%%%%%%%%%%%%%%%%%
 \section{Einstein--Maxwell Solutions: the Formalism}
%%%%%%%%%%%%%%%%%%%%%%%%%%%%%%%%%%%%%%%%%%%%%%%%%%%%%%%%%%%%%%%%%%%

When taking into account the backreaction on the geometry,
the solutions  above should result in EM  solitons and BHs
approaching at infinity the AdS$_D$ background.  
To construct them, 
we start with the following action principle in $D$-spacetime dimensions
(with $D=2k+3$ again):
\begin{eqnarray}
\label{action}
I_0=\frac{1}{16 \pi }\int_{\mathcal{M}} d^D x \sqrt{-g} (R-2 \Lambda-F^2)
-\frac{1}{8 \pi}\int_{\partial\mathcal{M}} d^{D-1} 
x\sqrt{-\gamma}K,
\end{eqnarray}
where % $G_D$ is the gravitational constant in $D$ dimensions, 
$\Lambda=-(D-1)(D-2)/(2 L^2)$ is the cosmological constant 
and $F=dB$ is the electromagnetic field strength. 
Here, $\mathcal{M}$ is 
a $D$-dimensional manifold with metric $g_{\mu \nu }$, and 
$K$ is the trace of the extrinsic 
curvature $K_{ab}=-\gamma_{a}^{c}\nabla_{c}n_{b}$ of 
the boundary $\partial M$ with unit normal $n^{a}$ and induced metric $\gamma_{ab}$.

As usual, the classical equations of motion are derived by setting 
the variations of the action (\ref{action}) to zero,
which results in the  
 EM system:
\begin{eqnarray}
R_{\mu\nu}-\frac{1}{2}R g_{\mu\nu}=\frac{(D-1)(D-2)}{2L^2}g_{\mu\nu}+2~T_{\mu\nu},~~~
\nabla_{\mu}F^{\mu\nu}=0.
\label{EMeqs}
\end{eqnarray}
Here, 
$T_{\mu\nu}=\left(F_{\mu\sigma}F^{\sigma}_{\nu}-
\frac{1}{4}F_{\sigma\rho}F^{\sigma\rho}g_{\mu\nu}\right) $
is the stress tensor of the electromagnetic field.

%%%%%%%%%%%%%%%%%%%%%%%%%%%%%%%%%%%%%%%%%%%%%%%%%%%%%%%%%%%%%%%%%%%
 \subsection{The ansatz and Equations}
%%%%%%%%%%%%%%%%%%%%%%%%%%%%%%%%%%%%%%%%%%%%%%%%%%%%%%%%%%%%%%%%%%%
The U(1) ansatz is still given by Equation (\ref{U1}), in terms of one magnetic potential $a_\varphi(r)$, only.
However,  taking into account the backreaction will deform the sphere $S^{2k+1}$,
with different factors for the two parts in Equation (\ref{s1}),
while the $g_{rr}$ and $g_{tt}$
metric functions will also receive corrections.
Then, after fixing the metric gauge, 
a sufficiently general metric ansatz contains three different functions depending on $r$ only,
with a line element

\vspace{-10pt}
 {\fontsize{9}{9}\selectfont
 \begin{eqnarray}
   \label{metric} 
 ds^2=\frac{1}{f(r)}
\left [
m(r) \left(\frac{dr^2}{N(r)}
+  r^2 d\Sigma_{k}^2 
      \right)
+n(r)r^2  \big(d\psi+ {\cal A} \big)^2
\right] -f(r)N(r) dt^2~,~~{\rm where}~~N(r)=1+\frac{r^2}{L^2}.
\end{eqnarray}}
 \textls[-19]{One can show that this ansatz is consistent, 
and the  Einstein equations
result in three second-order ODEs: }

\vspace{-10pt}
 {\fontsize{8.5}{8.5}\selectfont\begin{eqnarray} 
\nonumber
f''&-&\frac{(D-3)}{4}f
\left[
\frac{3f'^2}{f^2}+\frac{(D-4)}{(D-2)}\frac{m'^2}{m^2}
\right]  
+\frac{f'}{r} 
\left[
2(D-3)-\frac{(D-7)}{N}\frac{r^2}{L^2}
\right]
\\
\nonumber
&-&\frac{1}{D-2}\frac{f}{rN}
\left[
\left((D-2)\frac{r^2}{L^2}+(D-3)^2 \right)\frac{m'}{m} 
\right.
\left. +(D-3)\frac{n'}{n}
\right]
\\
\nonumber
&+&\frac{D-3}{D-2}f
\left[
\frac{f'n'}{fn}-\frac{m'n'}{2mn}+\frac{(D-4)(2D-5)}{2(D-3)}\frac{f'm'}{fm}
+\frac{1}{r^2N} \left(
1+\frac{(D-1)(D-2)}{(D-3)}\frac{r^2}{L^2}-\frac{n}{m}
\right)
\right]
\\
&-&\frac{(D-1)m}{L^2N}
-\frac{2f^2}{(D-2)r^2}
\left[
\frac{a_\varphi'^2}{n}+\frac{6(D-3)}{r^2Nm}a_\varphi^2
\right]
=0,
\label{ecf} 
%\end{eqnarray}
%
%
%
%\begin{eqnarray}
\\
\nonumber
m''&-&m 
\left[
\frac{D}{2(D-2)}\frac{m'^2}{m^2}
+\frac{D-5}{2}\frac{f'^2}{f^2}
+\frac{2(D-1)m}{L^2Nf}
\right]
+\frac{m}{r} \left[
 \frac{f'}{f}\left(2+\frac{D-6}{N}\right)
-\frac{D-4}{(D-2)N}\frac{n'}{n}
\right]
\\
\nonumber
\label{ecm}
%&&
%{~~~~~~~~}
&+&\frac{m'}{(D-2)rN}
\left[
2(D-3)+(D-2)^2\frac{r^2}{L^2}
\right]
+\frac{(D-4)m}{(D-2)}
\left(
\frac{f'n'}{fn}+\frac{D-4}{2}\frac{f'm'}{fm}-\frac{m'n'}{2mn}
\right)
\\
%\nonumber
%&&
%{~~~~~~~~}
&+&2(D-1)\frac{m}{r^2N}\left(N-\frac{D-1}{D-2}\right)
+\frac{2(D-1)n}{(D-2)r^2N}
-\frac{4f}{(D-2)r^2}
\left[
\frac{m a_\varphi'^2}{n}+\frac{2(D-5)a_\varphi^2}{r^2N}
\right]=0~,
\end{eqnarray}}

 \vspace{-10pt}
 {\fontsize{8}{8}\selectfont\begin{eqnarray}
\nonumber
%&&
n''&-& \frac{1}{2}n
 \left[
\frac{n'^2}{n^2}+(D-5)\frac{f'^2}{f^2}+\frac{(D-3)(D-4)}{(D-2)}\frac{m'^2}{m^2}
\right]+\frac{n'}{(D-2)r}
\left[
D(D-2)-\frac{(3D-8)}{N}
\right]
\\
\nonumber
&+&\frac{2(D-1)n}{(D-2)r^2N}
\left[
(D-2)N-1-(D-3)\frac{n}{m}-(D-2)\frac{m}{f}\frac{r^2}{L^2}
\right]
+\frac{4(D-3)f}{(D-2)r^2}
\left(
a_\varphi'^2-\frac{6n}{r^2Nm}a_\varphi^2
\right)
\\
\nonumber
&+&\frac{n}{(D-2)}
\left[
 (D(D-8)+14)\frac{m'n'}{2mn}
-\frac{(D-4)^2}{2}\frac{f'n'}{fn}
+(D-3)(D-4)\frac{f'm'}{fm}
\right]
\\
\label{ecn}
%&&
&+&\frac{n}{r}
\left[
\left(2+\frac{D-6}{N}\right)\frac{f'}{f}
-\left(
2+\frac{D(D-8)+14}{(D-2)N}
\right)\frac{m'}{m}
\right]
=0 \
,
\end{eqnarray} }together with a 
first-order constraint  

\vspace{-10pt}
 {\fontsize{8}{8}\selectfont
 \begin{eqnarray}
\label{constr}
\nonumber
&&
 \frac{1}{2}r^2 Nm 
\left[
(D-3)(D-4)\frac{f'm'}{fm} 
+(D-4)\frac{f'n'}{fn}
-(D-3)\frac{ m'n'}{mn}
\right]
\\
&&
\nonumber
+ m \left[   D-3+(D-1)(D-2)\frac{r^2}{L^2}\left(\frac{m}{f}-1\right) \right]-(D-3)n  
+ 2f \left[\frac{Nm}{n}a_\varphi'^2-2(D-3)\frac{a_\varphi^2}{r^2} \right]
\\
&&
\nonumber
-(D-3)rm'
\left[
D-3+(D-2)\frac{r^2}{L^2}+\frac{(D-4) }{4}\frac{Nrm'}{m}
\right]
-\frac{rm n'}{n}\left[D-3+(D-2)\frac{r^2}{L^2} \right]
\\
&&
%\nonumber
+(D-2)\frac{r mf'}{f}
\left[
D-4+(D-3)\frac{r^2}{L^2} -\frac{(D-5)}{4}\frac{N rf'}{ f}
\right]
=0,
\end{eqnarray} }which is a differential consequence of Equations (\ref{ecf}), (\ref{ecm}) and (\ref{ecn}).
The Maxwell equations imply that the magnetic potential solves the following 2nd order ODE

\vspace{-10pt}
 {\fontsize{8}{8}\selectfont 
\begin{eqnarray}
\label{ecA}
&& 
 a_\varphi''+a_\varphi'
\left[
\frac{D-4}{r}+\frac{2r}{NL^2}+ \frac{(D-4) m'}{2m}-\frac{(D-6)f'}{2f}-\frac{n'}{2n}
\right]
-\frac{2(D-3)n}{r^2 Nm}a_\varphi=0.
\end{eqnarray}}
\unskip

%%%%%%%%%%%%%%%%%%%%%%%%%%%%%%%%%%%%%%%%%%%%%%%%%%%%%%%%%%%%%%%%%%%
 \subsection{The Asymptotics}
%%%%%%%%%%%%%%%%%%%%%%%%%%%%%%%%%%%%%%%%%%%%%%%%%%%%%%%%%%%%%%%%%%%

Unfortunately, it seems that
no analytical techniques can be used  
to construct  in closed form
the solutions of the above equations. 
%Footnote is not preferred. We refer to it in the main text or you can refer it in other places of the main text. Please confirm if this is okay.
Although we mention 
the existence for $D=3$ of an exact solution
of EM Equations (\ref{EMeqs})
with a line element
%\begin{eqnarray}
%\nonumber
$ 
ds^2=\frac{dr^2}{N(r)U(r)}+\frac{U(r)}{(1+c_m^2)^2}r^2d\varphi^2-N(r) dt^2,
%\end{eqnarray} 
$
where
%\begin{eqnarray}
%\nonumber 
$
U(r)=1+c_m^2\frac{L^2}{r^2}\log N(r)
$ and $N=1+r^2/L^2$.
The  U(1) potential is
%\begin{eqnarray}
%\nonumber 
$
B=\frac{c_m L}{2(1+c_m^2)}\log N(r)d\varphi
$,
 diverging at infinity.
This solution has been discussed in 
\cite{Clement:1993kc, Hirschmann:1995he, Cataldo:1996ue, Dias:2002ps, Cataldo:2004uw},
and is usually interpreted as a magnetic soliton.
To our knowledge, its BH generalizations have not 
yet been considered in the literature.

However, one can construct an approximate expression
valid for large-$r$ and also another one close to $r=0$ or event horizon.
In constructing the far field form of the solutions,
we assume that the boundary spacetime topology  
is the product of time and  a round sphere $S^{D-2}$,
while the magnetic potential approaches a constant value.
Then, a direct computation leads to the following 
 large-$r$  expansion of the solutions:

\vspace{-10pt}
 {\fontsize{8.5}{8.5}\selectfont
 \begin{eqnarray}
\label{f-inf}
f(r)&=&1-\left( \frac{34}{25}\delta_{D,7}+\frac{5}{7}\delta_{D,9}  \right) \frac{c_m^2}{L^2}\left(\frac{L}{r} \right)^4
+ \frac{159}{49}\delta_{D,9}  \frac{c_m^2}{L^2}\left(\frac{L}{r} \right)^6 
\\
&&
\nonumber
+\bigg[
\frac{\hat \alpha}{L^{D-1}}+
          \frac{c_m^2}{L^2}\bigg(\frac{12}{9}\delta_{D,5}-\frac{652}{105}\delta_{D,7}+\left(\frac{214}{21}-\frac{4}{9}\frac{c_m^2}{L^2}\right) \delta_{D,9} \bigg)\log\left(\frac{L}{r}\right)
\bigg]
\left(\frac{L}{r} \right)^{D-1} +\dots,
%\end{eqnarray}
%
\\
%\begin{eqnarray}
\label{m-inf}
m(r)&=&1-\left( \frac{18}{25}\delta_{D,7}+\frac{3}{7}\delta_{D,9}  \right) \frac{c_m^2}{L^2}\left(\frac{L}{r} \right)^4
+ \frac{73}{49}\delta_{D,9}  \frac{c_m^2}{L^2}\left(\frac{L}{r} \right)^6 
\\
&&
\nonumber
+\bigg [
\frac{\hat \beta}{L^{D-1}}+
          \frac{c_m^2}{L^2}\bigg(\frac{4}{5}\delta_{D,5}-\frac{20}{21}\delta_{D,7}+\left(\frac{2}{3}-\frac{20}{63}\frac{c_m^2}{L^2}\right) \delta_{D,9} \bigg)\log\left(\frac{L}{r}\right)
\bigg]
\left(\frac{L}{r} \right)^{D-1} +\dots,
%\end{eqnarray}
%
\\
%\begin{eqnarray}
\label{n-inf}
n(r)&=&1-\left( \frac{68}{25}\delta_{D,7}+\frac{10}{7}\delta_{D,9}  \right) \frac{c_m^2}{L^2}\left(\frac{L}{r} \right)^4
+ \frac{514}{49}\delta_{D,9}  \frac{c_m^2}{L^2}\left(\frac{L}{r} \right)^6
\\
&&
\nonumber
+\bigg[
\frac{(D-2)(\hat \alpha-\hat \beta)}{L^{D-1}}
+\frac{c_m^2}{L^2}\bigg(\frac{4}{15}\delta_{D,5}-\frac{244}{105}\delta_{D,7}+\frac{2}{441}\left(1155- \frac{82c_m^2}{L^2}\right) \delta_{D,9} \bigg)
\\
&&
\nonumber
+
          \frac{c_m^2}{L^2}\bigg(\frac{24}{5}\delta_{D,5}-\frac{184}{7}\delta_{D,7}+8\left(\frac{75}{9}- \frac{c_m^2}{9L^2}\right) \delta_{D,9} \bigg)
					\log\left(\frac{L}{r}\right)
\bigg ] 
\left(\frac{L}{r} 
\right)^{D-1} +\dots,
\end{eqnarray} }together with
 \begin{eqnarray}
\nonumber
a_\varphi(r)&=& c_m-
\left(
2\delta_{D,7}+\frac{3}{2}\delta_{D,9}
\right)c_m
\left(\frac{L}{r}\right)^{2}
+3c_m\delta_{D,9}
\left(\frac{L}{r}\right)^{4}
\\
%&&
\label{a-inf}
&+&(D-3)c_m
	\log\left(\frac{L}{r}\right)\left(\frac{L}{r} 
\right)^{D-3} 
+\mu \left(\frac{1}{r}\right)^{D-3}
+\dots,
\end{eqnarray} 
valid for $D=5$, $7$, and $9$.
The corresponding expression becomes more complicated for higher $D$, with no general
pattern becoming apparent.
For any value of $D$, terms of higher order in $L/r$ depend on the two 
constants $\hat \alpha$ and $\hat \beta$ and also on the magnetic parameters $c_m$, $\mu$. 
In addition, one can verify that the asymptotic metric is still (AdS) maximally symmetric, i.e.,
 to leading order,
the
Riemann tensor is
$R_{\mu \nu}^{~~\lambda \sigma}=-(\delta_\mu^\lambda \delta_\nu^\sigma
-\delta_\mu^\sigma \delta_\nu^\lambda)/L^2$.

The corresponding expansion
near the origin $r=0$ reads
\begin{eqnarray}
\label{zero}
&&
f(r) = f_0 + f_2 r^2 +O(r^4), 
~~
m(r) =m_0+m_2 r^2+O(r^4),  
\\
\nonumber
&&
n(r) = m_0+ m_2r^2 +O(r^4), ~~ 
a_{\varphi}(r) = u r^2+ a_4 r^4+ +O(r^6),
\end{eqnarray}
with  
\begin{eqnarray} 
&&
f_2=\frac{m_0-f_0}{L^2}+\frac{4f_0^2 u^2}{(D-2)m_0},~~
m_2= \frac{4f_0 u^2}{2D-7}-\frac{3(D-2)(f_0-m_0)m_0}{(2D-7)f_0L^2}-\frac{3n_2}{2D-7}~, 
\\
\nonumber
&&
a_4=\frac{u}{L^2}\left[1+\frac{L^2n_2}{m_0}-\frac{2Dm_0}{(D-1)f_0}-\frac{16 u^2f_0L^2}{(D-2)(D+1)m_0} \right],
\end{eqnarray}
and 
 $
\{
m_0,f_0,u,n_2 
\}$
free parameters, with $f_0>0$ and $m_0>0$.
This implies that the solitons should be viewed as deformations of the 
AdS background, both parts in the $S^{D-3}$  metric sector shrinking to zero as $r\to 0$,
while $g_{tt}$ and $g_{rr}$ stay finite and nonzero.
 
There are also BH solutions. They possess an event horizon located at $r=r_H$,
an approximate form of the solution there being
 \begin{eqnarray}
\nonumber
&&
f(r) = f_2 (r-r_H)^2+ O\left(r-r_H\right)^3,~~ 
m(r)= m_2 (r-r_H)^2 + O\left(r-r_H\right)^3, 
\\
\label{near-horizon}
&&
n(r) = n_2 (r-r_H)^2  
+ O\left(r-r_H\right)^3, 
~~
a_{\varphi}(r) =  a_{\varphi }^{(0)}+ O\left(r-r_H\right)^2,
\end{eqnarray}
with 
 $
\{
f_2,m_2,n_2,a_{\varphi }^{(0)}
\}$
undetermined parameters.

%%%%%%%%%%%%%%%%%%%%%%%%%%%%%%%%%%%%%%%%%%%%%%%%%%%%%%%%%%%%%%%%%%%
 \subsection{The Mass Computation}
%%%%%%%%%%%%%%%%%%%%%%%%%%%%%%%%%%%%%%%%%%%%%%%%%%%%%%%%%%%%%%%%%%%

When evaluating  the  mass of the solutions for the far field expressions (\ref{f-inf})--(\ref{n-inf}),
one finds a divergent expression even in the absence  of a Maxwell field. 
The general remedy for this situation is to add counterterms, 
i.e., coordinate invariant functionals of the 
intrinsic boundary geometry that are specifically designed to cancel 
out the divergences \cite{Balasubramanian:1999re}. 
This procedure has the advantage of being intrinsic to the spacetime of interest, 
and it is unambiguous once
the counterterm action is specified.
Thus, we have to supplement the action %in \hl{Label} 
(\ref{action}) with  
(see \cite{Balasubramanian:1999re, Das:2000cu, Emparan:1999pm,Skenderis:2000in}):

 \vspace{-10pt}
 {\fontsize{8.5}{8.5}\selectfont\begin{eqnarray}
I_{\mathrm{ct}}^{(0)} &=&\frac{1}{8\pi }\int d^{D-1}x\sqrt{-\gamma 
}\left\{ -\frac{D-2}{\ell }-\frac{L \mathsf{\Theta }\left( D-4\right) 
}{2(D-3)}\mathsf{R}-\frac{D ^{3}\mathsf{\Theta }\left( D-6\right) 
}{2(D-3)^{2}(D-5)}\left[\mathsf{R}_{ab}\mathsf{R}^{ab}-
\frac{D-1}{4(D-2)}\mathsf{R}^{2}\right] 
\right.
\nonumber  
\\
\label{Lagrangianct} 
&&+\frac{L ^{5}\mathsf{\Theta }\left( D-8\right) 
}{(D-3)^{3}(D-5)(D-7)}\left[ 
\frac{3D-1}{4(D-2)}\mathsf{RR}^{ab}\mathsf{R}_{ab}
-\frac{D^2-1}{16(D-2)^{2}}\mathsf{R}^{3}\right. 
 \nonumber \\
&&\left. -2\mathsf{R}^{ab}\mathsf{R}^{cd}\mathsf{R}_{acbd}\left. 
-\frac{D-1}{4(D-2)}\nabla _{a}\mathsf{R}\nabla ^{a}\mathsf{R}+\nabla 
^{c}\mathsf{R}^{ab}\nabla _{c}\mathsf{R}_{ab}\right] +...\right\} ,
\end{eqnarray} }where $\mathsf{R}$ and $\mathsf{R}_{ab}$ are the curvature 
and the Ricci tensor associated with the induced metric $\gamma $. 
In this series, new terms
 enter at every new even value of $D$, as denoted by the 
step-function ($\mathsf{\Theta }\left( x\right) =1$
 provided $x\geq 0$, and vanishes otherwise).

However, in the presence of matter fields, additional
counterterms may be needed to regulate the mass and action of solutions
%Please refer to Ref. 28 in the main text.
\cite{Taylor:2000xw},
a situation which is not unusual in AdS physics. 
This is the case for the EM
solitons and BHs discussed in this paper.
The  supplementary
counterterm  has a rather complicated form,
with two different terms:
%Footnote is not preferred. We refer to it in the main text or you can refer it in other places of the main text. Please confirm if this is okay.
 \begin{eqnarray}
\label{IM} 
I_{\mathrm{ct}}^{(M)} =
\frac{1}{8\pi }\int d^{D-1}x\sqrt{-\gamma }
\bigg \{
\log\left(\frac{L}{r}\right) \mathsf{T}_0+\mathsf{T}_1\bigg \},
 \end{eqnarray}
where
 \begin{eqnarray}
&&
\nonumber
\mathsf{T}_0=
-\frac{L}{2}{\rm F}^2\delta_{D,5}
+
\frac{L^3}{20}\mathsf{R}{\rm F}^2\delta_{D,7}
-
\frac{L^5}{336}{\rm F}^2\left(\frac{2}{7}\mathsf{R}^2-\frac{1}{L^2}{\rm F}^2\right)\delta_{D,9}
+\dots,
\\
\label{T0T1}
&&
\mathsf{T}_1=  
\frac{L }{4}c_1(D){\rm F}^2\mathsf{\Theta }\left( D-6\right) 
+
\frac{L^3}{8}c_2(D)\mathsf{R}{\rm F}^2\mathsf{\Theta }
\left( D-8\right) 
+\dots~,  
\end{eqnarray}
with $c_1(D)=\delta_{D,7}+1/2 \delta_{D,9}+\dots$ and $c_2(D)=-\delta_{D,9}/21+\dots$.
In addition, ${\rm F}=c_m d{\cal A}$ is the electromagnetic tensor 
induced on the boundary by the bulk gauge field. 

A general expression of the Maxwell counterterm 
has been proposed in 
 \cite{Taylor:2000xw},
which,
 for $D=7, 9$, contains few other terms,
both in $\mathsf{T}_0$  and $\mathsf{T}_1$
(note that the extra terms possess
 the same leading order behaviour as those in (\ref{T0T1})).
However, we have found that the  counterterms
 in \cite{Taylor:2000xw} 
fail to regularize the action of the solutions in this work.
The problem seems to reside in the expression of some overall $D-$dependent 
coefficients there.
After fixing the value of those coefficients,
the results coincide with those displayed in this work

Using these counterterms, 
one can construct a boundary stress 
tensor from the total action $I=I_0+I_{\mathrm{ct}}^{(0)}+ I_{\mathrm{ct}}^{(M)}$ 
by defining
\begin{eqnarray}
\label{stress}
T^{ab}
=\frac 2{\sqrt{-\gamma}}\frac{\delta I  }{\delta\gamma_{ab}}~.
\end{eqnarray}%
Then, a conserved charge 
associated with a Killing vector $\xi ^{a}$ at infinity can be calculated using the~relationship: 
\begin{equation}
{\mathfrak Q}_{\xi }=\oint_{\Sigma }d^{D-2}S^a \xi ^{b}T_{ab},
\label{Mcons}
\end{equation}%
where $\Sigma $ is the sphere at infinity. 
The conserved mass/energy $M$ is the charge associated 
with the time translation symmetry, with $\xi =\partial /\partial t$. 
	
This prescription results in a finite expression of the mass,
which contains three different terms
\begin{eqnarray}
\label{MT} 
M=M_{(0)}+M_{(c)}+M_{(m)}, 
\end{eqnarray}  
with
\begin{eqnarray}
\label{M0} 
M_{(0)}= -\frac{{  \Omega}_{D-2}}{16\pi} \frac{(D-2)\hat \alpha+ \hat \beta}{ L^{2}} ,
\end{eqnarray}
 a standard term fixed by the constants 
$\hat \alpha, \hat \beta$ that enter the far field expansion.
There is also the usual Casimir  term $M_{(c)}$ that occurs for odd 
dimensions  \cite{Emparan:1999pm} 
 \begin{eqnarray}
\label{Mc} 
M_{(c)}= \frac{{  \Omega}_{D-2}}{8\pi} \frac{(D-2)!!^2}{(D-1)!}L^{D-3},
\end{eqnarray} 
with e.g.,
 \begin{eqnarray}
\label{Mcn} 
M_{(c)}=\frac{3 \Omega_{3}}{64\pi}L^2~~{\rm for}~~D=5,~~
M_{(c)}=-\frac{5 \Omega_{5}}{128\pi}L^4~~{\rm for}~~D=7,~~
M_{(c)}=\frac{35 \Omega_{7}}{1024\pi}L^6~~{\rm for}~~D=9.
\end{eqnarray} 
In addition, there is also a nontrivial contribution from the magnetic field
\begin{eqnarray}
\label{Mm} 
M_{(m)}= \frac{{ \Omega}_{D-2}}{16\pi} c_m^2 L^{D-5}
\left[
\frac{4}{15}\delta_{D,5}+\frac{458}{21}\delta_{D,7}
+\left(-\frac{4807}{84}+\frac{1285 c_m^2}{441 L^2}\right)\delta_{D,9}+\dots
\right]~.
\end{eqnarray}
Note that, in the above expression, ${ \Omega}_{D-2}$ is the total area of the angular sector.

%%%%%%%%%%%%%%%%%%%%%%%%%%%%%%%%%%%%%%%%%%%%%%%%%%%%%%%%%%%%%%%%%%
\subsection{Other Quantities}
%%%%%%%%%%%%%%%%%%%%%%%%%%%%%%%%%%%%%%%%%%%%%%%%%%%%%%%%%%%%%%%%%%
In addition, the BHs also possess some quantities determined 
by the horizon data in (\ref{near-horizon}).
The~Hawking temperature $T_H$ can be computed by evaluating the surface gravity
or by demanding regularity of the Euclideanized
manifold as $r\to r_H$.
This results in: 
\begin{eqnarray}
T_H=
\frac{1}{2\pi}\left(1+\frac{r_H^2}{L^2} \right)\frac{f_2}{\sqrt{m_2}}~.
\label{temp}
\end{eqnarray}
The horizon is a deformed $S^{D-2}$-sphere, with a line element
\begin{eqnarray}
\label{horizon-metric}
d \sigma_H^2=\frac{r_H^2}{f_2}\left(  
m_2 d\Sigma_k^2+n_2\big(d\psi+ {\cal A} \big)^2
\right),
\end{eqnarray}
its area $A_H$  being
\begin{equation}
\label{hor_area}
A_{\rm H} = r^{D-2}_{H}{  \Omega}_{D-2}
\sqrt{\frac{m^{D-3}_2 n_2}{f^{D-2}_2}}~.
\end{equation}
In addition, to have a measure of the squashing of the horizon, 
we introduce the deformation parameter 
\begin{eqnarray}
\label{epsilon}
\epsilon=\frac{n(r)}{m(r)}\bigg|_{r=r_H}=\frac{n_2}{m_2},
\end{eqnarray} 
which gives the ratio of the two parts  
parts in Equation (\ref{horizon-metric}).
 
These static Lorentzian solutions also make the Euclidean action extreme as the analytic
continuation in time has no effect at the level of the equations of motion.
Then, the tree level Euclidean action $I$ of these 
solutions can be evaluated by integrating the Killing 
identity $\nabla^\mu\nabla_\nu  \zeta_\mu=R_{\nu \mu}\zeta^\mu,$
for~the Killing vector $\zeta^\mu=\delta^\mu_t$, together with the  Einstein 
equation $R_t^t={(R - 2\Lambda-F^2)/2}$.
 In this way, one can isolate
 the bulk action
contribution at infinity and at $r=0$ (or $r=r_H$).
As usual, the surface integral term at infinity contains divergences
that are  
canceled by the Gibbons--Hawking term in the action~(\ref{action}) together with the counterterms
(\ref{Lagrangianct}) and (\ref{IM}).
The final expression for the total Euclideanized action is found in terms of boundary data  
at infinity and also for BHs at the horizon. 
For solitons, one finds $I=\beta M$ (with $\beta$ an arbitrary periodicity of  the Euclidean time).
For BHs, one finds $I=\beta M-S$
(this~time with $\beta=1/T_H$),
which implies an entropy of solutions,
 as computed from the Gibbs--Duhem relation,
$S=A_H/4$, as expected.

%%%%%%%%%%%%%%%%%%%%%%%%%%%%%%%%%%%%%%%%%%%%%%%%%%%%%%%%%%%%%%%%%%%
 \section{Einstein--Maxwell Solutions: The Results}
%%%%%%%%%%%%%%%%%%%%%%%%%%%%%%%%%%%%%%%%%%%%%%%%%%%%%%%%%%%%%%%%%%%

Although an analytic or approximate solution 
of the Equations (\ref{ecf})--(\ref{ecA})
 appears to be intractable, we 
present arguments here for the existence of nontrivial solutions,
which smoothly interpolate between the 
asymptotic expansions (\ref{f-inf})--(\ref{a-inf}) and the origin expansion (\ref{zero}) or 
the horizon expansion~(\ref{near-horizon}).  
Both solitons and BHs are found for $D=5$, $7$, and $9$,
by adapting the numerical techniques previously used  for rotating EM--Chern--Simons solutions
\cite{Blazquez-Salcedo:2015kja, Blazquez-Salcedo:2016rkj}.

The system of four non-linear coupled ODEs
for the functions $(f, m, n; a_\varphi)$ with appropriate boundary conditions 
(which follow
straightforwardly from Relations (\ref{f-inf})--(\ref{a-inf}) and Relation (\ref{zero})), 
was solved by using the software package COLSYS developed
by Ascher, Christiansen and Russell \cite{COLSYS,COLSYS2}. 
This solver uses a collocation method for the boundary conditions and an adaptive mesh selection procedure.
In the numerics, we employ a compactified radial coordinate $x$ 
(with $x=1-r_H/r$ for BHs and $x=r/(1+r)$ in the case of solitons, such that $0\leq x\leq 1$).
The solutions generated in this way have a typical relative precision of $10^{-8}$ or better, 
with around $10^3$ points in the mesh.

In the numerics, we fix the scale factor by taking a value $L=1$
for the AdS length scale. 
In~addition, to simplify the picture,
we did not include the value of the corresponding Casimir terms in the curves for the mass.
Since the equations of the model  are invariant under the change
of sign of  $a_\varphi$, we consider positive values of the magnetic parameter only, $c_m>0$.

\begin{figure}[H]
    \centering
    
    \begin{tabular}{cc}
   \includegraphics[width=50mm,scale=0.75,angle=-90]{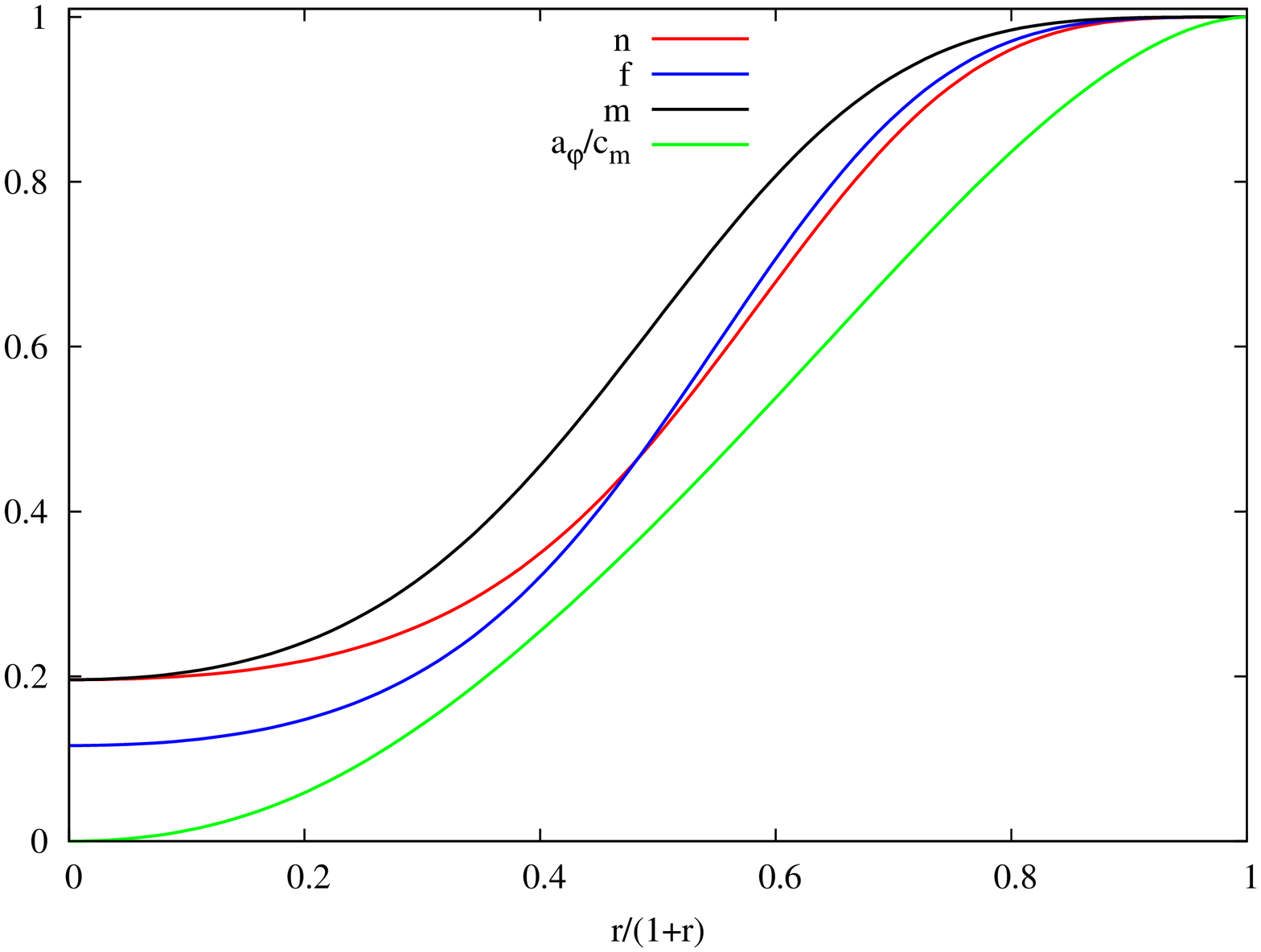}
& \includegraphics[width=50mm,scale=0.75,angle=-90]{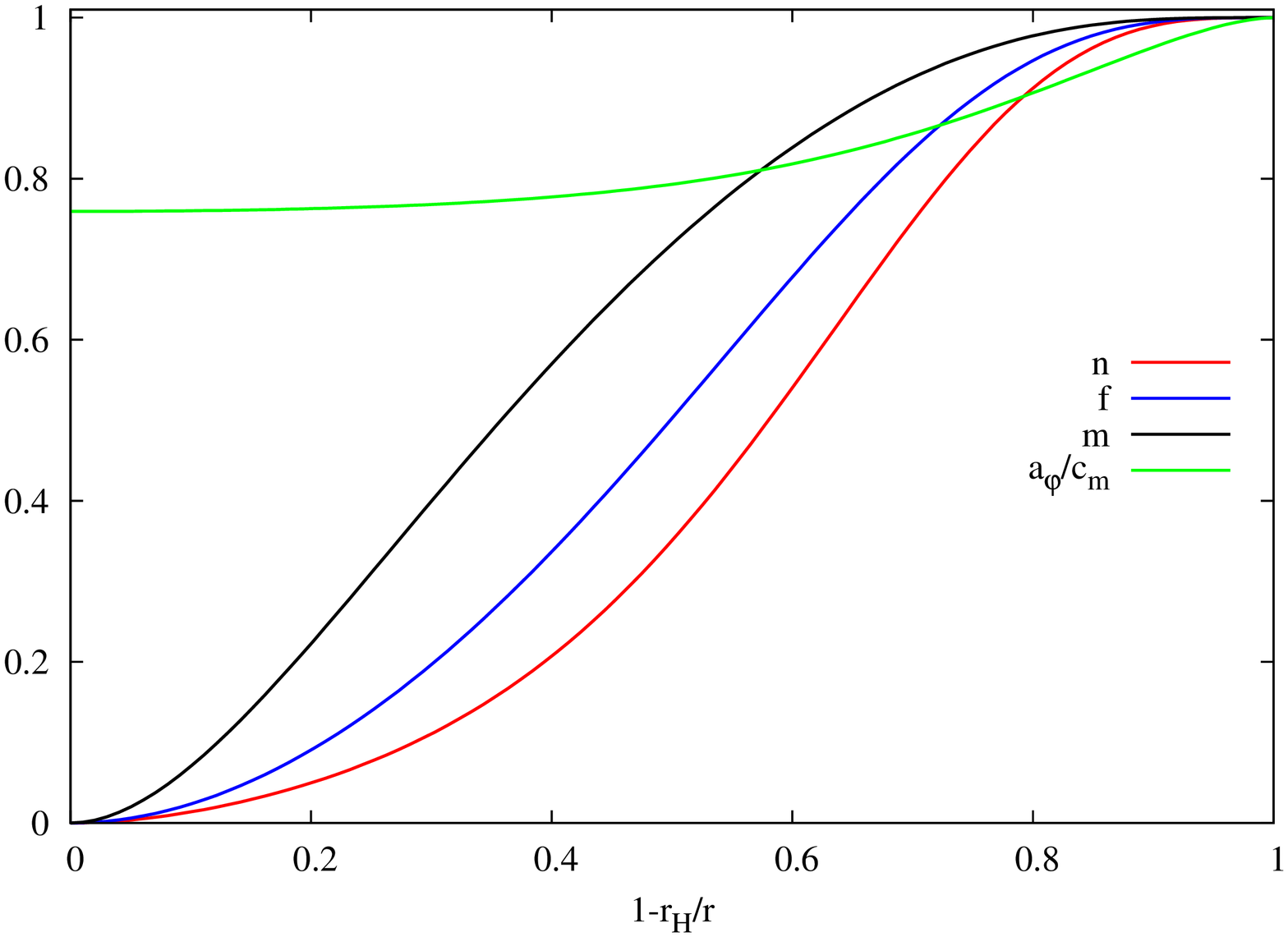}\\
({\bf a})&({\bf b})\\
\end{tabular}

%    \begin{subfigure}[b]{0.45\textwidth}
%        \includegraphics[width=50mm,scale=0.75,angle=-90]{soliton.eps}
%         \caption{}
%        \label{fig:ak_excitations}
%    \end{subfigure}~~~~
%    \begin{subfigure}[b]{0.45\textwidth}
%        \includegraphics[width=50mm,scale=0.75,angle=-90]{BH.eps}
%        \caption{}
%        \label{fig:omega_excitations}
%    \end{subfigure}
 
		    \caption{(\textbf{a}) the profiles of a typical soliton with $c_m=2$ and (\textbf{b}) a typical black hole with $r_H=1$ and 
				$c_m=4$. Both profiles correspond to $D=5$ and $L=1$.
				}
		\label{profiles}		
\end{figure}

For all the solutions we studied, the metric functions $f(r)$, $m(r)$, $n(r)$ 
and the magnetic potential $a_\varphi(r)$
interpolate
monotonically between the corresponding values at $r=0$
(or $r = r_H$) and the asymptotic values at
infinity, without presenting any local extrema. 
A typical example of solutions is shown in Figure \ref{profiles}
for a soliton (Figure \ref{profiles}a) and a BH (Figure \ref{profiles}b).

 %%%%%%%%%%%%%%%%%%%%%%%%%%%
 %%%%%%%----------NEW Figure  1------------%%%%%%

%%%%%%%----------NEW Figure  1-----------%%%%%%

%%%%%%%----------NEW Figure  4------------%%%%%%

%%%%%%%%%%%%%%%%%%%%%%%%%%%%%%%%%%%%%%%Figure 4
%\begin{figure}[H]
%    \centering
%    
%    \begin{tabular}{cc}
%   \includegraphics[width=50mm,scale=0.75,angle=-90]{soliton.eps}
%& \includegraphics[width=50mm,scale=0.75,angle=-90]{BH.eps}\\
%({\bf a})&({\bf a})\\
%\end{tabular}
% 
%
%		
%%    \begin{subfigure}[b]{0.45\textwidth}
%%        \includegraphics[width=50mm,scale=0.75,angle=-90]{soliton.eps}
%%         \caption{}
%%        \label{fig:ak_excitations}
%%    \end{subfigure}~~~~
%%    \begin{subfigure}[b]{0.45\textwidth}
%%        \includegraphics[width=50mm,scale=0.75,angle=-90]{BH.eps}
%%        \caption{}
%%        \label{fig:omega_excitations}
%%    \end{subfigure}
% 
%		    \caption{(\textbf{a}) the profiles of a typical soliton with $c_m=2$ and (\textbf{b}) a typical black hole with $r_H=1$ and 
%				$c_m=4$. Both profiles correspond to $D=5$ and $L=1$.
%				}
%		\label{profiles}		
%\end{figure}
%%%%%%%%%%%%%%%%%%%%%%%%%%%%%%%%%%%%%%Figure 4

%%%%%%%%%%%%%%%%%%%%%%%%%%%%%%%%%%%%%%%%%%%%%%%%%%%%%%%%%%%%%%%%%%
\subsection{The Solitons}
%%%%%%%%%%%%%%%%%%%%%%%%%%%%%%%%%%%%%%%%%%%%%%%%%%%%%%%%%%%%%%%%%%
  
The solitons are fully characterized by the value of the parameter $c_m$,
which enters the large$-r$ expansion of
the magnetic potential.
 
The numerical results for
$D=5,7,9$
indicate the existence of one single branch of solutions only. 
This branch starts at $c_m=0$  (which corresponds to vacuum $AdS_D$)
 and extends continuously up to some limiting value,  $c_m=c_{m}^*$. 
The limiting value $c_m^*$ depends on the dimension of the spacetime and $L$; 
for $D=5,7,9$, it satisfies with very good accuracy ($0.05\%$)
the linear relation
\begin{eqnarray}
 c_{m}^* =  (1.241+0.2055 D )L. 
\end{eqnarray}

The solutions with $0 \le c_m < c_m^*$ are regular everywhere. 
However,  the limit $c_m\to c_{m}^*$, is singular,
with
both the Ricci and the Kretschmann scalars diverging at the origin, $r=0$. 
In addition, we could not find regular solitons with $c_m>c_m^*$. 
Thus, we conclude that these EM-AdS solitons cannot exist
for arbitrarily large values of the magnetic field on the boundary.

\begin{figure}[H]
    \centering
		
		\begin{tabular}{cc}
\includegraphics[width=70mm,scale=0.8,angle=0]{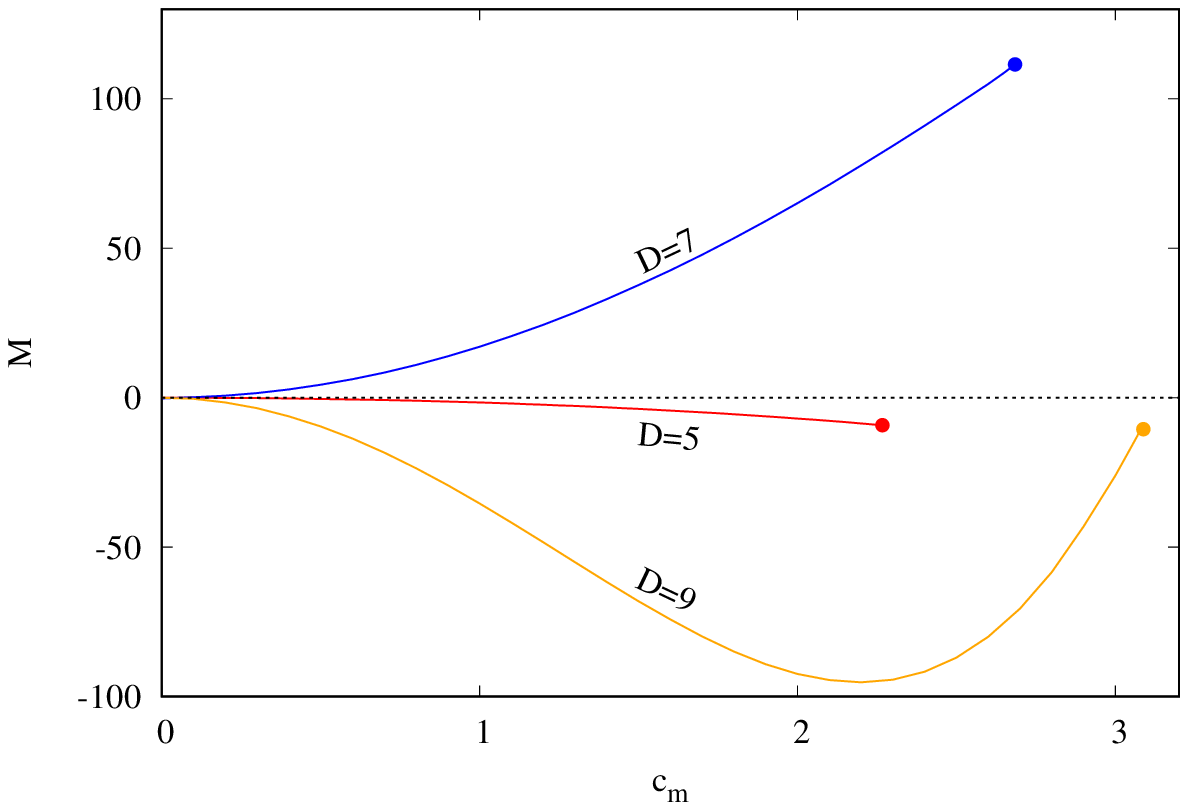}
&     \includegraphics[width=70mm,scale=0.8,angle=0]{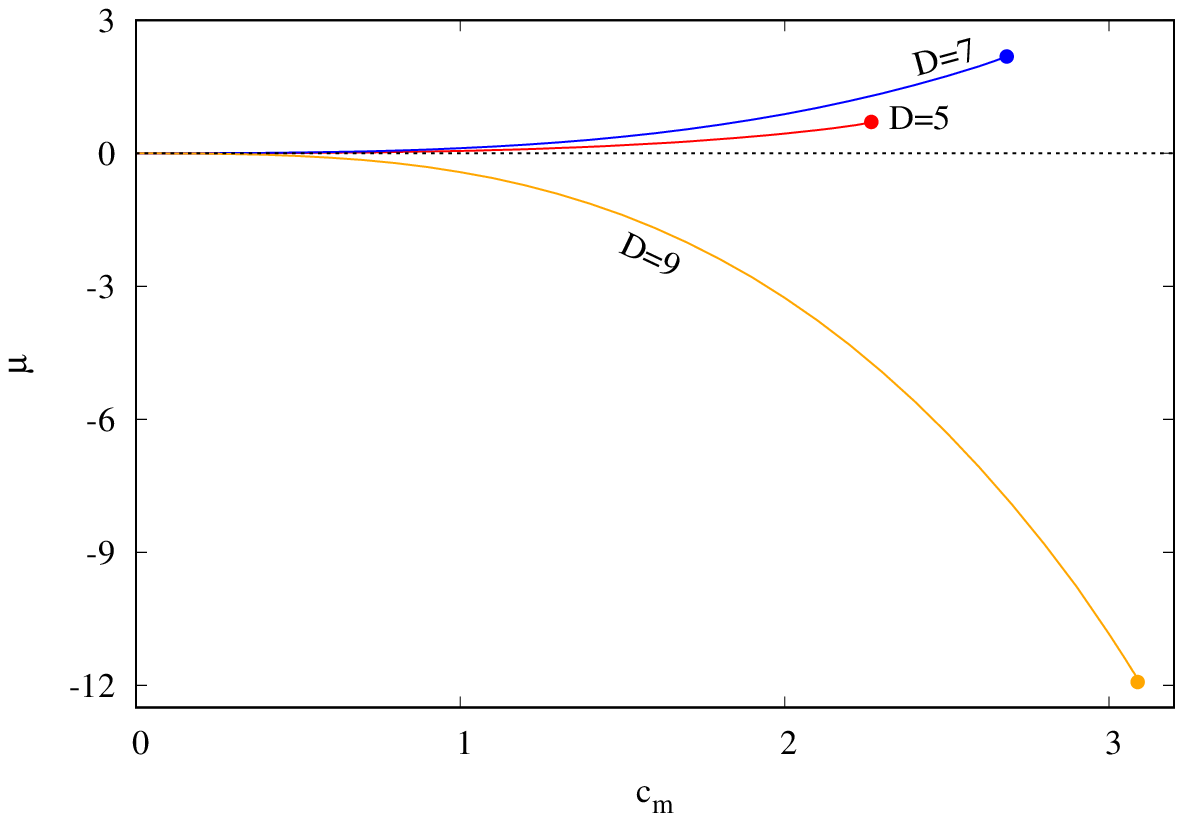}\\
({\bf a})&({\bf b})\\
\end{tabular}
	
%    \begin{subfigure}[b]{0.45\textwidth}
%        \includegraphics[width=70mm,scale=0.8,angle=0]{plot_mass_cm_D_2.eps}
%        \caption{}
%        \label{fig:mass_solitons_D}
%    \end{subfigure}
%    \begin{subfigure}[b]{0.45\textwidth}
%        \includegraphics[width=70mm,scale=0.8,angle=0]{plot_mu_cm_D.eps}
%        \caption{}
%        \label{fig:mu_solitons_D}
%    \end{subfigure}
		
    \caption{(\textbf{a}) Mass $M$ vs. $c_m$ and (\textbf{b}) $\mu$ vs. $c_m$ for static 
		solitons in $D=5$ (\textbf{red}), $D=7$ (\textbf{blue}) and $D=9$ (\textbf{orange}). 
		In both figures, the dots mark the endpoints of the branch of solitons.}
		
		\label{fig:solitons_D}
		
\end{figure}

In Figure \ref{fig:solitons_D}a, we show the mass $M$ vs. the parameter $c_m$, 
the dots marking the position of the limit solutions at $c_{m}^*$.
One can notice that the $M(c_m)$ curve depends on the value of the spacetime dimension.
However, the mass remains finite as  $c_m\to c_{m}^*$.
 
In Figure \ref{fig:solitons_D}b, we show a similar 
plot for the magnetic moment $\mu$, and again for  $D=5$, $7$, and~$9$. 
One can see that, as expected, the magnitude of the magnetic moment always increases with $c_m$  
(note~also that the sign of $\mu$ is negative in $D=9$). 	

%%%%%%%%%%%%%%%%%NEW Figure 2's position
%%%%%%%----------NEW Figure  2------------%%%%%%
\subsection{The Black Holes}
%%%%%%%%%%%%%%%%%%%%%%%%%%%%%%%%%%%%%%%%%%%%%%%%%%%%%%%%%%%%%%%%%%
As expected, these solitons possess BH generalizations. 
In what follows, we will focus on the $D=5$ case, 
although the  described   generic properties are expected to hold for any $D=2k+1$.

The magnetized BHs are described by two independent parameters, 
a natural choice being $c_m$ and the mass $M$. 
Then, the solutions can be generated in two ways:
(i) by adding a horizon inside the solitons described in the previous Section, or
(ii) by introducing a magnetic field into the SAdS BH (which has $c_m=0$).
  
One important difference with respect to the solitonic case is that,  for BHs,
the parameter $c_m$ is no longer bounded.  
However, some basic properties of the solutions
now depend on $c_m$ being smaller or larger than the critical value $c_m^*$
noticed above for solitons
 (for $D=5$, one finds $c_m^* \simeq 2.269$ for $L=1$).
 
\begin{figure}[H]
    \centering

    \begin{tabular}{cc}
\includegraphics[width=70mm,scale=0.8,angle=0]{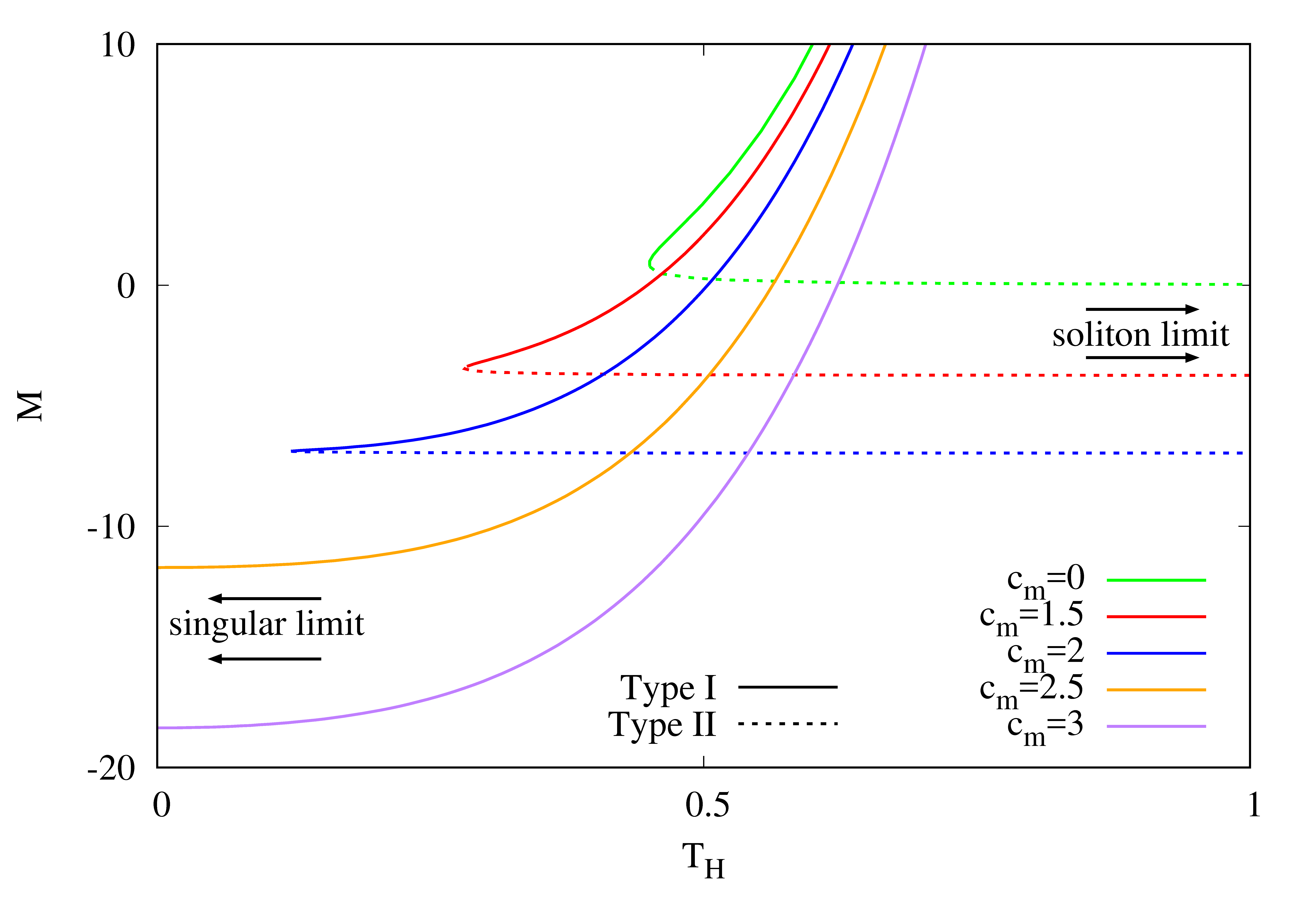}
&\includegraphics[width=70mm,scale=0.8,angle=0]{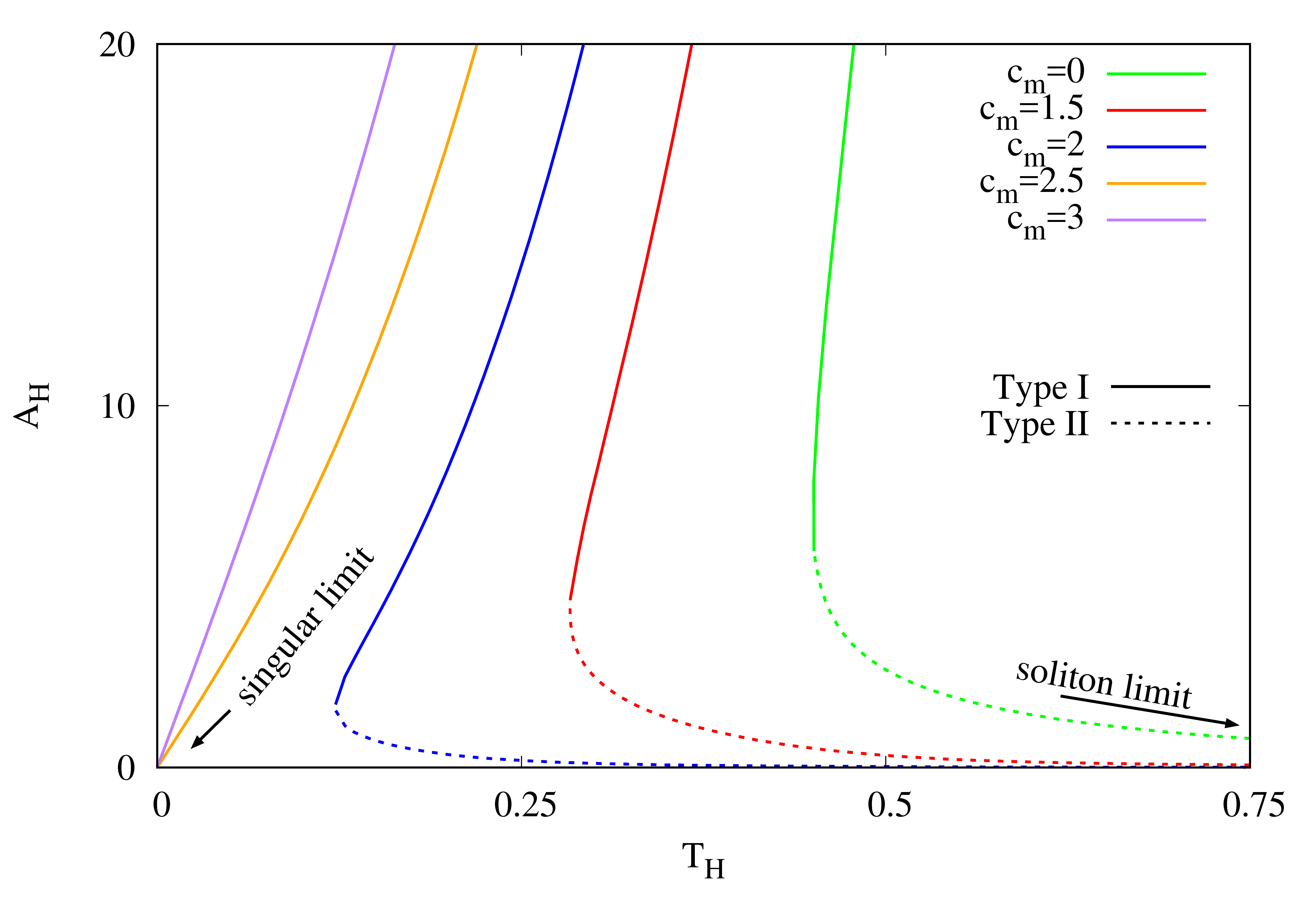}\\
({\bf a})&({\bf b})\\
\end{tabular}

%    \begin{subfigure}[b]{0.45\textwidth}
%        \includegraphics[width=70mm,scale=0.8,angle=0]{plot_mass_T_2.jpg}
%        \caption{}
%        \label{fig:mass_T_bhs}
%    \end{subfigure}
%    \begin{subfigure}[b]{0.45\textwidth}
%        \includegraphics[width=70mm,scale=0.8,angle=0]{plot_Ah_T_2.jpg}
%        \caption{}
%        \label{fig:Ah_T_bhs}
%    \end{subfigure}
		    \caption{(\textbf{a}) the mass $M$ is shown vs. temperature $T_H$ for static  $D=5$
		black holes with several different values of the magnetic parameter $c_m$.
For $c_m < c_m^*$, one finds both type I solutions (continuous lines) and type II solutions (dashed lines). 
Type II black holes can be deformed into solitons in the limit $T_H \rightarrow \infty$. 
For $c_m>c_m^*$, only Type I black holes are present and 
 the limit $T_H \rightarrow 0$ is singular; 
(\textbf{b})~the~horizon area $A_H$ is shown vs. temperature  $T_H$
for the same solutions.
 Note that for type II black holes ($c_m<c_m^*$),
 the horizon shrinks to zero as $T_H \rightarrow \infty$, 
a limit which corresponds to a soliton deformation of the AdS background. 
For $c_m>c_m^*$, only Type I solutions are found, while the limit $T_H \rightarrow 0$ has $A_H \rightarrow 0$,
 being singular.}
				\label{fig:bhs_1}
	\end{figure}

Let us start by considering families of configurations with fixed values of $c_m$ and varying
the temperature.
In Figure \ref{fig:bhs_1}a, we show
 the mass $M$ vs. 
the temperature $T_H$ for magnetized BHs with several values of $c_m$. 
A similar plot is shown in Figure \ref{fig:bhs_1}b for the horizon area $A_H$ vs. the temperature $T_H$.  
One can see that, for $0 \le c_m < c_m^*$, the
BHs possess
 two distinct branches,
which we shall call $type~I$ and $type~II$. 
In the vacuum limit, these correspond to the large and the small branches, respectively, of~SAdS BHs.
For type I BHs (solid lines), 
 the mass and the horizon area increase with the temperature.
For type II BHs (dashed lines), the mass and the horizon area decrease as the temperature increases. 
This set of BHs is especially interesting, since they can be deformed continuously into solitons, 
as the horizon size tends to zero.

Similar to the vacuum SAdS case, these solutions exist above a minimal value of temperature only.  
This minimal temperature  decreases with increasing $c_m$, reaching zero as $c_m=c_m^*$.
In fact, when $c_m \ge c_m^*$, the type II branch completely disappears.
Consider for example, the curves for $c_m=2.5$ (orange solid line) and $c_m=3$ (purple solid line) 
in Figure \ref{fig:bhs_1}a,b . 
One can see that only type I BHs are found in these cases, 
and their temperature can reach $T_H=0$. 
However, contrary to what happens for the
 electrically charged Reissner--Nordstr\"om-AdS BHs, the extremal limit does not correspond to a regular configuration.
 Although the mass approaches a finite value there, 
the horizon area vanishes. 
In~fact, the configurations with $T_H=0$ 
possess divergent Ricci and Kretschmann scalars at the horizon. 

\begin{figure}[H]
    \centering

    \begin{tabular}{cc}
\includegraphics[width=70mm,scale=0.8,angle=0]{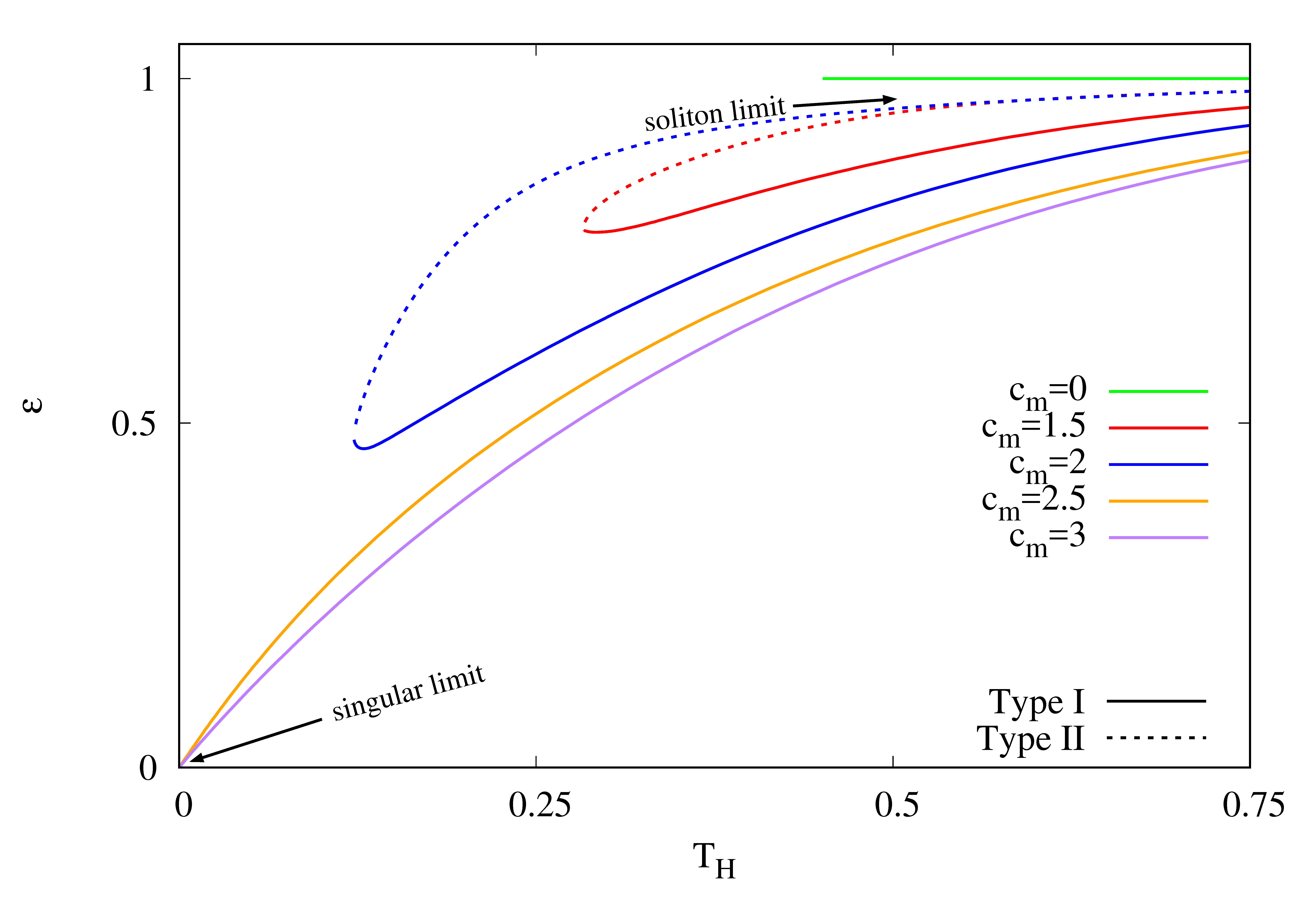}
&\includegraphics[width=70mm,scale=0.8,angle=0]{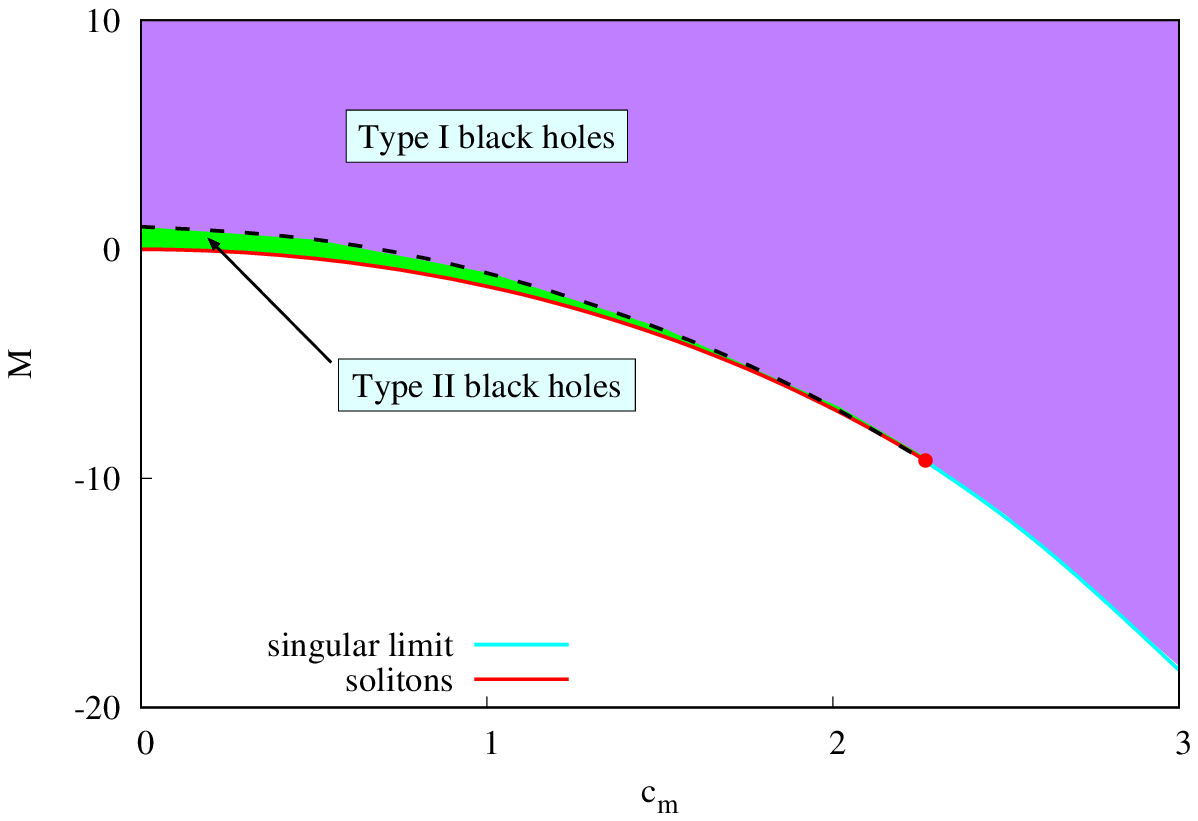}\\
({\bf a})&({\bf b})\\
\end{tabular}
		
%    \begin{subfigure}[b]{0.45\textwidth}
%        \includegraphics[width=70mm,scale=0.8,angle=0]{plot_eps_T_2.jpg}
%        \caption{}
%        \label{fig:eps_T_bhs}
%    \end{subfigure}
%    \begin{subfigure}[b]{0.45\textwidth}
%        \includegraphics[width=70mm,scale=0.8,angle=0]{plot_mass_cm_domain_2.eps}
%        \caption{}
%        \label{fig:mass_cm_domain}
%    \end{subfigure}
		
    \caption{(\textbf{a}) Deformation parameter $\epsilon$ vs. $T_H$ for static black holes in $D=5$ for different values of $c_m$ in various colors. Type I solutions are plotted with continuous lines and type II with dashed lines. Note that, in type II BHs ($c_m<c_m^*$), the horizon becomes spherical as it shrinks when $T_H \rightarrow \infty$, forming a soliton. For $c_m>c_m^*$, only Type I is present, and in the limit $T_H \rightarrow 0$ the solution becomes singular; (\textbf{b})~mass vs. $c_m$ for static solitons and black holes in $D=5$. In \textbf{red}, we plot the solitons, and the \textbf{red} dot marks the endpoint. The \textbf{blue} line marks the singular limit of black holes. The \textbf{black} dashed line separates the two types of static BHs: Type I (\textbf{purple} area) and Type II (\textbf{green} area). Type II black holes can be contracted to form a soliton.}
		
		\label{fig:bhs_2}
		
\end{figure}

The singular nature of these solutions can also
be appreciated in Figure \ref{fig:bhs_2}a, 
where we show the deformation parameter $\epsilon$ 
vs. the parameter $c_m$. 
First, one should notice that $\epsilon<1$ for all solutions;
thus, a magnetic field increases the relative size of the round $S^{D-3}$
part of the horizon as compared to that of the corresponding $S^1$ part. 
In addition, one can see that $\epsilon \to 0$ as $T_H\to 0$,
which likely indicates a change in the topology of the horizon, from spherical to planar.  
 
%%%%%%%NEW Figure 3's position
%%%%%%%----------NEW Figure 3------------%%%%%%
%%%%%%%%%%%%%%%%%%%%%%%%%%%Figure 2
%%%%%%%%%%%%%%%%%%%%%%%%%%%% Figure 2
%%%%%%%----------NEW Figure 3------------%%%%%%
 \unskip

 %%%%%%%----------NEW Figure  4------------%%%%%%
 %%%%%%%%%%%%%%%%%%%%%%%%%%%Figure 3
%%%%%%%%%%%%%%%%%%%%%%%%%%%% Figure 3
 %%%%%%%----------NEW Figure  4------------%%%%%%

 \medskip
Finally,  in Figure \ref{fig:bhs_2}b, we show the full domain of existence of the 
$D=5$
magnetized BHs, 
in a mass $M$ vs. $c_m$ diagram. 
This domain is bounded by three different sets of solutions. 
First, at $c_m=0$, one finds the standard  SAdS BHs, with $0 \le M < \infty$.
For $0 \le c_m < c_m^*$, the lower bound of the domain of existence is formed by the set of solitons, 
represented by a red line in Figure \ref{fig:bhs_2}b. 
The red dot marks the endpoint of the soliton branch (note that the solitons  always possess a lower mass than 
the BHs with the same $c_m$).
However, the BHs exist also for 
$c_m \ge c_m^*$, in which case the lower bound of the domain of existence is no longer given by solitons.
Instead, one finds that if one decreases the mass of the BHs, 
one reaches the {\it singular limit} described in the previous paragraphs. 
This set of singular solutions is plotted as a blue line in Figure \ref{fig:bhs_2}b.  
 
Then, the magnetized BHs are found in the shaded area,
 which represent regular configurations with different temperatures. 
The type I BHs fill the purple area and are always above the configurations with minimum temperatures 
(that set is marked in Figure \ref{fig:bhs_2}b with a dashed black line). 
The type II BHs connect with the soliton limit and fill the green area between the solitons (red line) 
and the minimum temperature configurations (dashed black line).

%%%%%%%%%%%%%%%%%%%%%%%%%%%%%%%%%%%%%%%%%%%%%%%%%%%%%%%%%%%%%%%%%%
\section{Conclusions}
%%%%%%%%%%%%%%%%%%%%%%%%%%%%%%%%%%%%%%%%%%%%%%%%%%%%%%%%%%%%%%%%%%

Some recent results in the literature 
\cite{Herdeiro:2015vaa, Costa:2015gol, Herdeiro:2016xnp, Herdeiro:2016plq}
indicate that, for $D=4$,
the  known solutions
of the EM system in a globally AdS$_4$ background 
 represent only `the tip of the iceberg'.
New solutions without Minkowski spacetime counterparts were shown to
exist (this includes EM solitons), 
being supported by the  confining box  behaviour of the AdS spacetime.

The main purpose of this work was to inquire about the possible existence of similar configurations
in more than four dimensions.
The solutions reported here are likely to be the simplest one can consider in this context.
For an odd number of spacetime dimensions, we use a suitable static metric ansatz 
and a purely $magnetic$ U(1) field,
which factorized the angular dependence.
As such, the problem reduced to solving a set of ODEs,
which significantly simplifies the setup as compared to the $D=4$ case.

Our numerical results in this paper cover the cases $D=5$, $7$, and $D=9$, although
similar solutions should exist for an arbitrary dimension $D=2k+1$.
Some basic results are similar to those known for $D=4$.
For example, the existence of solutions can be traced back
to the fact that, in contrast to the flat space, the Maxwell equations
in an AdS background possess solutions that are finite everywhere.
Moreover, these solutions possess self-gravitating generalizations, 
while a BH can be added at their~center.

However, some new features occur as well.
For example, for $D>4$, the mass of the solutions, as defined in the usual way, diverges, 
despite the spacetime being asymptotically AdS, and one has to supplement 
the boundary action with a matter counterterm.
Moreover, for gravitating solitons, one notices the existence of a maximal value of the
magnitude of the magnetic potential at infinity for~$D>4$. 

As avenues for future research,
we mention first the issue of electrically charged generalizations, which can be constructed
by supplementing  
 the gauge field ansatz with an electric potential,
\begin{eqnarray}
\label{U1e}
B= a_\varphi(r)  \big(d\psi+ {\cal A} \big)+a_0(r)dt~.
\end{eqnarray}
One should remark that,
although the problem remains codimension-1,
the presence of an electric field implies that the solutions necessarily rotate.

Purely electric, static and non-spherically symmetric solutions should exist as well,
supported by nontrivial asymptotics of the electric potential.
 However, they  would be less symmetric, with no cohomogeneity-1 ansatz  as in this work, and
 would be found as solutions of partial differential equations.
In fact, we have preliminary evidence for the existence of such configurations in $D=5$
dimensions.
They share some basic properties of their magnetic counterparts discussed here
(for~example, their mass as defined in the usual way, diverges logarithmically at infinity).

On a more conceptual level, 
it would be interesting to consider the  solutions in this work in an AdS/CFT context.
The fact that, for $D>4$, the EM system (subject to the symmetries in this work)
 does not correspond to a consistent truncation of a gauged supergravity model 
 %Footnote is not preferred. We refer to it in the main text or you can refer it in other places of the main text. Please confirm if this is okay.
makes it more difficult to obtain a CFT description. 
At the same time, Equation (\ref{action}) 
is the basic part of a gauged supergravity action.
Thus, we
expect some basic properties of the solutions in this work
 to also hold for 
generalizations within a supergravity framework.
If, for example, one supplements the $D=5$ action with a $U(1)$ Chern--Simons term, 
these static black holes are no longer solutions of the theory, unless one considers 
a planar horizon together with a special gauge field ansatz \cite{D'Hoker:2009mm}
(see also \cite{D'Hoker:2009bc,Ammon:2016szz}). However, it will certainly be important to study 
in the future possible extensions of these solutions to Einstein--Maxwell--Chern--Simons theory.

% \section{Conclusions} \hl{Please add}

\vspace{6pt}
%%%%%%%%%%%%%%%%%%%%%%%%%%%%%%%%%%%%%%%%%%%%%%%%%%%%%%%%%%%%%%%%%%%%%%%
\acknowledgments{We gratefully acknowledge support by
the DFG Research Training Group 1620 ``Models of Gravity''.
 E.R. acknowledges funding from the FCT-IF programme.
This work was also partially supported 
by  the  H2020-MSCA-RISE-2015 Grant No.  StronGrHEP-690904, 
and by the CIDMA project UID/MAT/04106/2013.  
J.L.B.-S. and J.K. gratefully acknowledge support by the grant FP7, Marie Curie Actions, People, International Research Staff Exchange Scheme (IRSES-606096).
F. N.-L. acknowledges funding from Complutense University under project PR26/16-20312.
 }

%%%%%%%%%%%%%%%%%%%%%%%%%%%%%%%%%%%%%%%%%%
\authorcontributions{%\hl{For research articles}
%Please add.
All authors contributed equally to this paper. All authors have read and approved the final manuscript.}
 
 %%%%%%%%%%%%%%%%%%%%%%%%%%%%%%%%%%%%%%%%%%
\conflictofinterests{The authors declare no conflict of interest.}

\bibliographystyle{mdpi}

%=====================================
% References, variant A: internal bibliography
%=====================================
\renewcommand\bibname{References}

\end{document}